\documentclass[aps,pre,groupedaddress,twocolumn,10pt, reprint,superscriptaddress, longbibliography]{revtex4-2}

\usepackage[utf8]{inputenc}
\usepackage{CJK}
\CJKfamily{gbsn}
\usepackage{graphicx,bm}
\usepackage{amsmath,amssymb,amsthm}
\usepackage{appendix}
\usepackage{array,booktabs}
\usepackage{pgfplots}
\usepackage{upgreek}
\usepackage{siunitx}[=v2]

\newcommand\pddiff[2]{\frac{\partial^2 #1}{\partial #2^2}}

\newcommand\ecoli{\textit{E. coli }}

\newcommand{\rr}{\bm{\mathrm{r}}}
\newcommand{\kk}{\bm{\mathrm{k}}}

\newcommand{\uu}{\bm{\mathrm{u}}}

\theoremstyle{definition}

\theoremstyle{remark}

\usepackage{tikz}
\usepackage{pgfplots}
\usepackage{color}
\usepackage[hidelinks]{hyperref}
\hypersetup{
    colorlinks,
    linkcolor={red!50!black},
    citecolor={blue!50!black},
    urlcolor={blue!80!black}
}
\usepackage[bf,FIGTOPCAP,nooneline]{subfigure}

\begin{document}

\begin{CJK*}{UTF8}{gbsn}
\title{Detecting active L\'evy particles using differential dynamic microscopy}

\author{Mingyang Li (李明洋)}
\affiliation{Center for Soft Condensed Matter Physics and Interdisciplinary Research \& School of Physical Science and Technology, Soochow University, 215006 Suzhou, China}

\author{Yu'an Li (李聿安)}
\email{yuan\_li@sjtu.edu.cn}
\affiliation{School of Physics and Astronomy, Shanghai Jiao Tong University, Shanghai 200240, China}
\affiliation{Institute of Natural Sciences and MOE-LSC, Shanghai Jiao Tong University, Shanghai 200240, China}

\author{H. P. Zhang (张何朋)}
\affiliation{School of Physics and Astronomy, Shanghai Jiao Tong University, Shanghai 200240, China}
\affiliation{Institute of Natural Sciences and MOE-LSC, Shanghai Jiao Tong University, Shanghai 200240, China}

\author{Yongfeng Zhao (赵永峰)}
\email{yfzhao2021@suda.edu.cn}
\affiliation{Center for Soft Condensed Matter Physics and Interdisciplinary Research \& School of Physical Science and Technology, Soochow University, 215006 Suzhou, China}

\date{\today}

\begin{abstract}
Detecting L\'evy flights of cells has been a challenging problem in experiments. The challenge lies in accessing data in spatiotemporal scales across orders of magnitude, which is necessary for reliably extracting a power-law scaling. Differential dynamic microscopy has been shown to be a powerful method that allows one to acquire statistics of cell motion across scales, which is a potentially versatile method for detecting L\'evy walks in biological systems. In this article, we extend the differential dynamic microscopy method to self-propelled L\'evy particles, whose run-time distribution has an algebraic tail. We validate our protocol using synthetic imaging data and show that a reliable detection of active L\'evy particles requires accessing length scales of an order of magnitude larger than its persistence length, if the variability in particle speed is moderate. Applying the protocol to experimental data of \ecoli and \textit{E. gracilis}, we find that \ecoli does not exhibit a signature of L\'evy walks, while \textit{E. gracilis} is better described as active L\'evy particles.
\end{abstract}

\maketitle 

\end{CJK*}

\section{Introduction}
Microorganisms exhibit various patterns of motility to navigate in a complex environment~\cite{Taktikos:2013}: Run-and-tumble motion of \textit{Escherichia coli}~\cite{Berg:1972} and \textit{Euglena gracilis}~\cite{tsang2018polygonal}, run-reverse-flip pattern of several marine bacteria~\cite{Taktikos:2013}, and run-reverse-wrap motion of \textit{Pseudomonas putida}~\cite{Alirezaeizanjani:2020}. Among these patterns, the run-and-tumble motion is arguably the simplest model that enables tactic behaviors of cells~\cite{Schnitzer:1993,Wadhams:2004}, and has attracted a lot of attention~\cite{Tailleur:2008,Angelani:2013,Cates:2013,santra2020run,datta_random_2024,loewe_anisotropic_2024}. The trajectory of a run-and-tumble motion is composed of straight "runs", interrupted by sudden changes of moving direction, which are referred to as "tumbles". In the simplest model considered in typical theoretical work~\cite{solon_active_2015}, the particle runs at constant speed and orientation. Tumbles happen randomly with a constant rate over time, which results in an exponential distribution of run time. Following the term in the literature, we use "run-and-tumble particles (RTP)" specifically for the model with exponentially distributed run time~\cite{solon_active_2015,kurzthaler2024characterization}.

However, experimental evidence suggests that some microorganisms and cells may exhibit L\'evy walks~\cite{bartumeus_helical_2003,harris_generalized_2012,ariel_swarming_2015,reynolds_current_2018,Figueroa:2018,figueroa-morales_e_2020,huo2021swimming,Junot:2022,li2025biased} on experimentally observed scales. For example, experiments on the molecular motor of \textit{E. coli} suggest that the run-time distribution has a power-law tail $t^{-\mu}$. The exponent $\mu$ is reported to be less than 3~\cite{cluzel2000ultrasensitive}, so that the mean squared displacement of the cell scales as $t^{4-\mu}$ on large time scales, which is superdiffusive. The tracking data of a flagellated alga \textit{E. gracilis} is also reported to exhibit a L\'evy walk~\cite{li2025biased}.

Nevertheless, detection of algebraic scaling from experimental data is a notoriously challenging problem~\cite{benhamou_how_2007,petrovskii_variation_2011}. Power laws suggest scale invariance. Their identification requires measurements over the order-of-magnitude variation in spatiotemporal scales, which is generally a challenging experimental task. Indeed, the consequences of the L\'evy walk in \textit{E. coli} have not been confirmed on large scales. A recent experimental characterization observes a diffusive regime~\cite{kurzthaler2024characterization} on a length scale of the order of 400 $\upmu$m, contradicting the expectation of a L\'evy walk.

Differential dynamic microscopy (DDM) is a high-throughput method that can simultaneously access scales across 2 orders of magnitude~\cite{Wilson:2011,Martinez:2012,kurzthaler2024characterization,zhao2024quantitative}, and thus has the potential to overcome experimental challenges. DDM measures the intermediate scattering function (ISF) of particles, defined as
\begin{equation}
    f(\kk,\tau)=\langle e^{-i\kk\cdot\Delta\rr(\tau)}\rangle\;.\label{eqn_isf_def}
\end{equation}
$\Delta\rr(\tau)$ is the displacement of the particle during the lag time $\tau$ in a steady trajectory, and $\kk$ is the wave vector in Fourier space. ISF is the probability density of particle displacement in Fourier space, and thus contains the full information on particle motion. Although ISF is defined from the trajectories of particles, it can be measured from the autocorrelation of density fluctuations of particles without the need to resolve particle trajectories. The field of view usually contains statistics of $10^4\sim 10^6$ particles. The advantage of acquiring high-throughput data across scales may help detect L\'evy walks.

In this article, we extend differential dynamic microscopy to detect active L\'evy particles (ALPs). Firstly, we use simulation-generated data to validate the spatial-temporal scales that allow distinguishing between an ALP and an RTP. Next, we analyze the experimental results of \textit{E. coli} (published in Ref.~\cite{kurzthaler2024characterization}) and \textit{E. gracilis}. We find that while the data of \textit{E. coli} are best fitted by an RTP model, the data of \textit{E. gracilis} exhibit a signature of active L\'evy particles on length scales from $10^1\ \upmu$m to $10^3\ \upmu$m.

The paper is organized as follows. We first introduce a paradigmatic model for ALPs and the renewal theory in Sec.~\ref{sec_model}. Then we show the key properties of the ISF of ALPs by exploiting a simplest case of ALPs in Sec.~\ref{sec_isf_alp}. In Sec.~\ref{sec_validation} we present a protocol for detecting ALPs using differential dynamic microscopy and validate using synthetic imaging data. In Sec.~\ref{sec_exp} we present analysis of experimental data of \ecoli and \textit{E. gracilis}. Finally, we summarize and discuss our results in Sec.~\ref{sec_summary}.

\section{Models}\label{sec_model}

\subsection{Run-and-tumble particles (RTP) and active L\'evy particles (ALP)}

\begin{figure}
    \centering
    \begin{tikzpicture}
        \begin{scope}[yshift=5.5cm]
            \node at (0,0) {\includegraphics[width=0.3\textwidth]{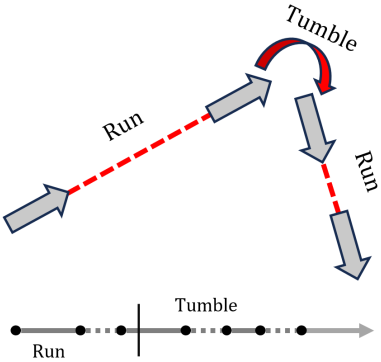}};
            \node at (-2.5,2) {(a)};
            \node at (-0.7,-2.7) {$t=0$};
            \node at (3.2,-2.1) {Time $t$};
            \node at (-0.4,-0.1) {$\mathbb{P}_R(\rr,\tau)$};
            \node at (2.8,1.5) {$\mathbb{P}_T(\rr,\tau)$};
            \node at (-1.95,-1.8) {$\varphi_R(\tau)$};
            \node at (0.3,-2.4) {$\varphi_T(\tau)$};
        \end{scope}
        
        \begin{scope}[xshift=-2.2cm]
            \node at (0,0) {\includegraphics{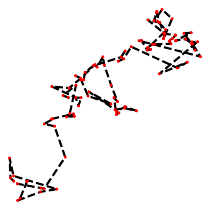}};
            \node at (-1.5,2) {(b)};
            \node at (0.7,-1) {$\varphi_R(\tau)\sim e^{-\tau}$};
        \end{scope}
        
        \begin{scope}[xshift=2.2cm]
            \node at (0,0) {\includegraphics{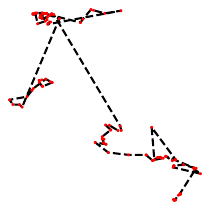}};
            \node at (-1.5,2) {(c)};
            \node at (0.7,1) {$\varphi_R(\tau)\sim \tau^{-\mu}$};
        \end{scope}
    \end{tikzpicture}
    
    \caption{\textbf{(a)} A paradigmatic model of run-and-tumble-like particles. A particle switches between run and tumble states. The probability density functions, or the propagators, that a running (tumbling) particle travels a distant $\rr$ after time $\tau$ are denoted as $\mathbb{P}_R(\rr,t)$ ($\mathbb{P}_T(\rr,t)$). The run and tumble time distributions are $\varphi_R(\tau)$ and $\varphi_T(\tau)$, respectively. \textbf{(b,c)} Typical trajectories of (b) a run-and-tumble particle and (c) an active L\'evy particle with exponent $\mu=2.5$. The RTP and the ALP have the same persistence length.}
    \label{fig:placeholder}
\end{figure}

The analysis of the intermediate scattering function (ISF) measured from experiments requires specifying a probing model to fit the data. We note that at this stage we cannot conclude \textit{a priori} whether the L\'evy walk of cells persists on infinitely large scales, and it is likely that there is a cut-off time beyond which cells undergo normal diffusion. The probing model, in principle, can take this large-scale cut-off time into account. However, a trade-off has to be considered that the probing model should include all the important ingredients but not overfit the data.

A good strategy for choosing a probing model is to start with the simplest possible model, and progressively include more ingredients if the simple model fails to fit the data. We thus first proceed without considering a large-scale cut-off time. If such a model fits the small-scale data well but misses large scales, we should consider a more complex probing model with large-time cut-off. We show in Sec.~\ref{sec_exp} that our experimental observations didn't reach this cut-off time. Thus, from now on, we focus our analysis on a simplest model of active L\'evy particles, and we stress that the detection of a power law only applies to the experimentally accessible scales.

We consider a paradigmatic model of run-and-tumble-like particles~\cite{kurzthaler2024characterization}. The particle switches between running and tumbling states. A running particle moves at a constant speed $v$ in a straight line and enters the tumbling state after a random run time $\tau$ with a probability distribution function (PDF) $\varphi_R(\tau)$. A tumbling particle diffuses passively and resumes running after a random tumble time $\tau'$ with a PDF $\varphi_T(\tau')$. When a particle resumes running, it randomly chooses a new direction $\uu$ uniformly distributed on a 2D or 3D unit sphere, and it randomly chooses a new swimming speed $v$ according to the Schultz distribution $P(v;\bar{v},\sigma_v)$ with mean speed $\bar{v}$ and standard deviation $\sigma_v$, which is defined as
\begin{equation}
    P(v;\bar{v},\sigma_v)=\frac{v^Z}{\Gamma(Z+1)}\left(\frac{Z+1}{\bar v}\right)^{Z+1}e^{-(Z+1)v/\bar v}\;, \label{eqn_schultz}
\end{equation}
and $Z=\bar v^2/\sigma_v^2-1$. The particle can be subjected to translational noise, with diffusion coefficient $D$, regardless of the state it enters.

The two models we consider differ in the asymptotic behavior of $\varphi_R(\tau)$ in $\tau\to\infty$. As in the literature~\cite{Tailleur:2008,solon_active_2015,zhao2024quantitative}, we refer to as run-and-tumble particles (RTPs) the model with exponentially distributed run and tumble time.
\begin{equation}
    \varphi_{R,T}^{\rm RTP}(\tau)=\frac{1}{\tau_{R,T}}\exp\left(-\frac{\tau}{\tau_{R,T}}\right)\;, \label{eqn_rtp_varphi}
\end{equation}
where $\tau_{R,T}$ are the mean run and tumble time, respectively. 

By contrast, $\varphi_R(\tau)$ of the active L\'evy particles (ALPs) have a power-law tail at large $\tau$. The power law must be truncated at small $\tau$ to ensure the normalization of $\varphi_R(\tau)$. For convenience, we use a Lomax distribution~\cite{lomax_business_1954,Zaburdaev:2015,figueroa-morales_e_2020,li2025biased} to introduce a continuous cutoff at small $\tau$.
\begin{equation}
    \varphi_{R}^{\rm ALP}(\tau)=\frac{\mu-1}{\tau_0}\left(1+\frac{\tau}{\tau_0}\right)^{-\mu}\;,\ \varphi_{T}^{\rm ALP}(\tau)=\frac{e^{-\tau/\tau_T}}{\tau_{T}}\;,\label{eqn_alp_varphi}
\end{equation}
where $\tau_0$ is the cutoff time scale. The mean run time is finite for $\mu>2$ and is given by
\begin{equation}
    \tau_R=\frac{\tau_0}{\mu-2}\;,\ \text{for }\mu>2\;.
\end{equation}
The choice of Lomax distribution has a simple microscopic interpretation that the rate of particle tumbling is $(\mu-1)/(\tau_r+\tau_0)$, where $\tau_r$ is the time since the beginning of the run. The time dependence in the rate indicates the non-Markovian nature of the ALP. 

\subsection{Renewal theory}
The ISF~\eqref{eqn_isf_def} of RTPs and ALPs can be calculated using renewal theory~\cite{Detcheverry:2015,zhao2024quantitative}. We briefly review the renewal theory for self-containing in this section. Renewal theory iteratively constructs the probability density functions (PDFs) for particles running or tumbling at a position $\rr$ and time $t$. It assumes that the time intervals between consecutive events (e.g., runs and tumbles for RTPs and ALPs) are statistically independent, such that the particle "renews" itself at each event. The PDF of an event occurring at a specific point $(\rr,t)$ in spacetime can then be constructed from the convolution of the probability of the last renewal event with the survival probability of the current state, which sequentially links to the initial distribution of the particle.

Specifically, the stochastic trajectory of a run-and-tumble-like particle is fully determined by the definition of run and tumble time distributions, $\varphi_{R,T}(\tau)$, and run and tumble propagators, $\mathbb{P}_{R,T}(\rr,\tau)$. The propagators $\mathbb{P}_{R,T}(\rr,\tau)$ measure the probability that a particle travels a distance $\rr$ during a time $\tau$ in a running or a tumbling state, respectively. The two models we considered share the same expressions of $\mathbb{P}_{R,T}(\rr,\tau)$, which are expressed in Fourier space as
\begin{align}
    \mathbb{P}_R^{\rm 3D}(k,\tau)=&\int_0^\infty P(v;\bar{v},\sigma_v)\exp(-Dk^2\tau)\frac{\sin(vk\tau)}{vk\tau}\,dv\;,  \\
    \mathbb{P}_R^{\rm 2D}(k,\tau)=&\int_0^\infty P(v;\bar{v},\sigma_v)\exp(-Dk^2\tau)J_0(vk\tau)\,dv\;, \label{eqn_swim_propagator_2d}  \\
    \mathbb{P}_T(k,\tau)=&\exp(-Dk^2\tau)\;.
\end{align}
We note that $\mathbb{P}_R$ depends on the spatial dimension and $J_0(x)$ is the 0th-order Bessel function.

Then we denote $P_R(\rr,\tau)$ and $P_T(\rr,\tau)$ as the probability densities of the particle displaced by a distance $\rr$ after a delay time $\tau$, conditioned on the particle in the running and tumbling state at time $\tau$, respectively. We denote $P_{R,T}(\kk,\tau)$ as their Fourier transforms. Since the system is isotropic, we drop the angular dependence, and the ISF can then be expressed as
\begin{equation}
    f_{\rm RT}(k,\tau)=P_R(k,\tau)+P_T(k,\tau)\;,\label{eqn_isf_rt}
\end{equation}
where $k=|\kk|$. Following Ref.~\cite{zhao2024quantitative}, $P_{R,T}(k,\tau)$ are given by a set of integral equations.
\begin{align}
P_R(k,\tau)&\!=\!P_R^0(k,\tau)\!+\!\int_0^\tau dt \, R(k, \tau\!-\!t)\varphi_R^0(t)\mathbb{P}_R(k,t)\;,\label{eq:renewalFT1}\\
R(k,\tau)&\!=\!R^1(k,\tau)\!+\!\int_0^\tau dt \, T(k,\tau\!-\!t)\varphi_T(t)\mathbb{P}_T(k,t)\;,\label{eq:renewalFT2} \\
P_T(k,\tau)&\!=\! P^0_T(k,\tau)\!+\!\int_0^\tau dt \, T(k,\tau\!-\!t)\varphi^0_T(t)\mathbb{P}_T(k,t)\;,\label{eq:renewalFT3}\\
T(k,\tau) &\!=\!T^1(k,\tau)\!+\!\int_0^\tau dt \, R(k,\tau\!-\!t)\varphi_R(t)\mathbb{P}_R(k,t)\;.\label{eq:renewalFT4}
\end{align}
$\varphi_{R,T}^0(\tau)=\int_\tau^\infty d\tau'\,\varphi_{R,T}(\tau')$ are the probabilities that the run or tumble time exceeds $t$, respectively. $R(\rr,\tau)$ and $T(\rr,\tau)$ are probabilities that a particle starts to run or tumble at displacement $\rr$ and lag time $\tau$, with $R(k,\tau)$ and $T(k,\tau)$ being their Fourier transforms. $P^0_{R,T}(\rr,\tau)$ is the probability that the particle reaches $\rr$ at time $\tau$ without having tumbled or run in $[0,\tau]$, respectively. We assume that the initial time and position of the particle is arbitrary in its steady trajectory, so that the fraction of running particles $p_R=\tau_R/(\tau_R+\tau_T)$. Then their Fourier transforms are given by
\begin{align}
P^0_{R}(k,\tau)&= p_{R} \mathbb{P}_{R}(k,\tau)\int_{\tau}^\infty dt \, \varphi_R(t)(t-\tau)/\tau_R \;, \\
P^0_{T}(k,\tau)&= (1-p_{R}) \mathbb{P}_{T}(k,\tau)\int_{\tau}^\infty dt \, \varphi_T(t)(t-\tau)/\tau_T \;.
\end{align}
$R^1(\rr,\tau)$ and $T^1(\rr,\tau)$ are the probabilities of starting the first run or tumble in displacement $\rr$ at time $\tau$, respectively. Their Fourier transforms are given by
\begin{align}
    R^1(k,\tau)&=(1-p_R)\mathbb{P}_T(k,\tau)\int_\tau^\infty dt\,\varphi_T(t)/\tau_T\;, \\
    T^1(k,\tau)&=p_R\mathbb{P}_R(k,\tau)\int_\tau^\infty dt\,\varphi_R(t)/\tau_R\;.
\end{align}

Then equations~\eqref{eq:renewalFT1}-\eqref{eq:renewalFT4} form a closed set, and $f_{\rm RT}(k,\tau)$ can be calculated numerically by time stepping. The Laplace transform of the ISF, $f_{\rm RT}(k,s)=\int_0^\infty f_{\rm RT}(k,\tau)\exp(-s\tau)\,d\tau$, can be solved analytically~\cite{zhao2024quantitative}. We note that the renewal theory is applied if the particle has finite run and tumble durations $\tau_{R,T}$, and is thus applied to ALPs with $\mu>2$.

\section{The intermediate scattering function of active L\'evy particles}\label{sec_isf_alp}

\begin{figure*}
    \centering
    \begin{tikzpicture}
    \def\hcolx{4.25cm}
    \def\hfigy{3.175cm}
        \begin{scope}[yshift=0.5*\hfigy]
            \node at (-1.5,1.7) {(a)};
            \node at (0,0) {\includegraphics{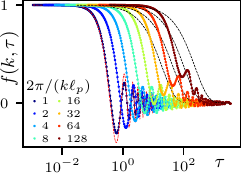}};
            \node at (-1,0.8) {$\mu=2.2$};
        \end{scope}
        
        \begin{scope}[yshift=0.5*\hfigy,xshift=\hcolx]
            \node at (-1.5,1.7) {(b)};
            \node at (0,0) {\includegraphics{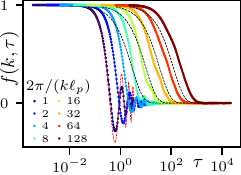}};   
            \node at (-1,0.8) {$\mu=2.8$};     
        \end{scope}
        
        \begin{scope}[yshift=-0.5*\hfigy]
            \node at (-1.5,1.7) {(c)};
            \node at (0,0) {\includegraphics{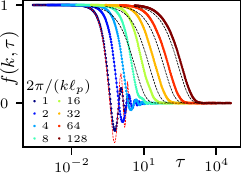}};  
            \node at (-1,0.8) {$\mu=3.5$};      
        \end{scope}
        
        \begin{scope}[yshift=-0.5*\hfigy,xshift=\hcolx]
            \node at (-1.5,1.7) {(d)};
            \node at (0,0) {\includegraphics{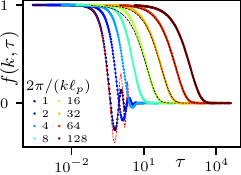}};   
            \node at (-1,0.8) {RTPs};     
        \end{scope}
        
        \begin{scope}[xshift=2.5*\hcolx]
            \node at (-3,3.3) {(e)};
            \node at (0,0) {\includegraphics{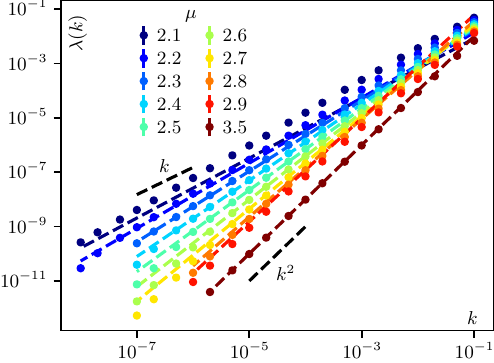}};        
        \end{scope}
    \end{tikzpicture}
    \caption{The intermediate scattering functions (ISFs) of active L\'evy particles (ALPs) and their asymptotic behavior in two dimensional space. We consider the simplest case $D=0$, $\sigma_v=0$, and $\tau_T=0$. \textbf{(a-d)} ISFs of ALPs (a-c) and RTPs (d) with fixed persistence length $\ell_p=1$ and varying wavenumber $k$. For ALPs, we use $\mu=2.2$ (a), 2.8 (b), 3.5 (c). Color encodes wavenumber $k$ normalized by the persistence length $\ell_p:=v_0\tau_R$. Circles represents ISFs measured from particle simulations using Eq.~\eqref{eqn_isf_def}, and solid lines shows the theoretical prediction calculated by numerically inverse Laplace transformation of Eq.~\eqref{eqn_fks_alp_simplest} and~\eqref{eqn_fks_rtp}. The red dashed lines show the swimming propagator~\eqref{eqn_swim_propagator_2d} in 2D with $D=0$ and $\sigma_v=0$ for $2\pi/k=\ell_p$. The black dashed lines shows the asymptotic function~\eqref{eqn_asym_alp} and~\eqref{eqn_asym_rtp} for $2\pi/k=16\ell_p$, $32\ell_p$, $64\ell_p$, $128\ell_p$. \textbf{(e)} The decaying rate $\lambda(k)$ of ISFs of ALPs as a function of $k$, where we fit the ISFs of ALPs by an exponential function $\exp(-\lambda(k)\tau)$. The dots represent $\lambda(k)$ measured from numerical ISFs calculated using Eq.~\eqref{eqn_isf_def}. Color encodes $\mu$. The dashed lines shows the asymptotic decaying rate shown in Eq.~\eqref{eqn_asym_alp}. We fix $\tau_0=1$ in this panel. Parameters: $v_0=1$.}
    \label{fig:ALP_asymptotic}
\end{figure*}

To analytically understand the key signatures of ALPs, we start from the simplest case where $D=0$ and $\tau_T=0$. We consider particles swimming at the same and constant speed $v_0$ so that $\sigma_v=0$. In Sec.~\ref{sec_ALP_ft} and~\ref{sec_sigmav} we will discuss the effect of a non-zero $\tau_T$ and $\sigma_v$. The simplest model reduces to a 2D or 3D L\'evy walk~\cite{Zaburdaev:2015,Zaburdaev:2016}. Using renewal theory, the intermediate scattering function (ISF) of ALPs in the Fourier-Laplace domain can be expressed as
\begin{equation}
    \tilde{f}(\tilde{k},\tilde{s})=\frac{(\mu-2)W(\tilde{s},\tilde{k};\mu-1)^2}{1-(\mu-1)W(\tilde{s},\tilde{k};\mu)}+W(\tilde{s},\tilde{k};\mu-2)    \label{eqn_fks_alp_simplest}
\end{equation}
where $\tilde{f}(\tilde{k},\tilde{s}):=f(k,s)/\tau_0$ is the dimensionless Laplace transform of the ISF. $s$ is the Laplace time, $\tilde{s}=s\tau_0$ and $\tilde{k}=kv_0\tau_0$ are dimensionless Laplace and Fourier variables, respectively. The dimensionless function $W(x,y;\mu)$ satisfies
\begin{equation}
W(s\tau_0,kv_0\tau_0;\mu)=\frac{1}{\mu-1}\mathcal{L}[\varphi_R(\tau)\mathbb{P}_R(k,\tau)](s)\;,
\end{equation}
with its explicit expression in 2D and 3D space, respectively.
\begin{align}
    W^{\rm 2D}(x,y;\nu)&=\int_0^\infty d t\,\frac{e^{-xt}J_0(yt)}{(1+t)^\nu}\;, \\
    W^{\rm 3D}(x,y;\nu)&=\int_0^\infty d t\,\frac{e^{-xt}\sin(yt)}{yt(1+t)^\nu}\; .
\end{align}
In Fig.~\ref{fig:ALP_asymptotic}a-c, we show examples of ISFs of ALPs with varying $\mu$ but fixed $\tau_R$ in real lag time $\tau$, where the numerical inverse Laplace transformation of Eq.~\eqref{eqn_fks_alp_simplest} is calculated. The ISFs of RTPs with the same $\tau_R$ are shown in Fig.~\ref{fig:ALP_asymptotic}d for comparison. Compared with RTPs, the ISFs of ALPs have more significant oscillations on large length scales. We will provide a detailed discussion in Sec.~\ref{sec_compare}.

\subsection{Asymptotic behavior in large scales}\label{sec_asymp}
Firstly, we analyze the asymptotic behavior of $\tilde{f}(\tilde{k},\tilde{s})$ on large scales. We note that $W(x,y;\mu)$ is not an analytic function at $x=y=0$, and the asymptotic behavior requires specifying the path of the limit $(x,y)\to (0,0)$~\cite{Schmiedeberg_2009}. We seek a following rescaling of the temporal and spatial coordinates to larger scales.
\begin{align}
    &\tilde{s} \mapsto b\tilde{s}\;, &\tilde{k} \mapsto b^\xi \tilde{k}\;, \label{eqn_asym_scale}
\end{align}
where $1>b>0$ is a scaling factor and $\xi>0$ is the exponent of spatial coordinates. Because $f(k,\tau)$ is a dimensionless function that evolves from 1 to 0 in $\tau\in[0,\infty)$, we seek an asymptotic function $g(k,\tau)$ that is invariant after rescaling $f(k,\tau)$. Then its Laplace transform $\tilde{f}$ should rescale as $\tilde{f}\mapsto b\tilde{f}$. The asymptotic form $g(\tilde{k},\tilde{s})$ of the Laplace-transformed ISF is then given by a proper choice of $\xi$ such that
\begin{align}
    g(\tilde{k},\tilde{s}):=\lim_{b\to 0} b\tilde{f}(b^\xi \tilde{k},b\tilde{s})  \label{eqn_def_asym}
\end{align}
has a non-trivial dependence on $s$ and $k$. 

We then expand $W(bx,b^\xi y;\nu)$ with respect to $b$ at finite $x,y>0$. The leading terms are given by
\begin{align}
    W(bx,b^\xi y;\nu)\sim  &\frac{1}{\nu-1}-\frac{xb}{(\nu-1)(\nu-2)}+A_\nu y^{\nu-1}b^{\xi(\nu-1)} \nonumber \\
    &-\frac{y^2b^{2\xi}}{d(\nu-1)(\nu-2)(\nu-3)}+\cdots\;,\label{eqn_w_asymp}
\end{align}
where $\cdots$ represents higher-order terms in $b$, and $A_\nu$ is a factor dependent on spatial dimension $d$.
\begin{align}
    A_\nu=\left\{\begin{array}{ll}
        \displaystyle\frac{\Gamma(1/2-\nu/2)}{2^\nu\Gamma(1/2+\nu/2)}\;, & d=2\;, \\[12pt] 
        \displaystyle\frac{\pi}{2\Gamma(\nu+1)\cos(\pi\nu/2)}\;, & d=3\;.\\
    \end{array}\right.
\end{align}

Next, we plug the expansion~\eqref{eqn_w_asymp} into Eq.~\eqref{eqn_fks_alp_simplest}. We denote
\begin{equation}
\tilde{f}(\tilde{k},\tilde{s})=\frac{N(\tilde{k},\tilde{s})}{D(\tilde{k},\tilde{s})}\;,\ D(\tilde{k},\tilde{s}):= 1-(\mu-1)W(\tilde{s},\tilde{k};\mu)\;. \label{eqn_fks_alp_ND}
\end{equation}
Letting $\nu=\mu$, $x=\tilde{s}$, and $y=\tilde{k}$ in Eq.~\eqref{eqn_w_asymp}, the leading terms in the denominator are
\begin{equation}
    D(\tilde{k},\tilde{s})\sim \frac{\tilde{s}b}{\mu-2}+\frac{\tilde{k}^2b^{2\xi}}{d(\mu-2)(\mu-3)}-(\mu-1)A_\mu\tilde{k}^{\mu-1}b^{\xi(\mu-1)}\;.\label{eqn_fks_deno_asymp}
\end{equation}
Similarly, the leading term in the numerator $N(\tilde{k},\tilde{s})$ is
\begin{equation}
    N(\tilde{k},\tilde{s})\sim \frac{1}{\mu-2}\;.\label{eqn_fks_nume_asymp}
\end{equation}
For $\lim_{b\to 0}bN/D$ to be finite and non-trivial, one needs non-vanishing $\tilde{k}$-dependent terms in Eq.~\eqref{eqn_fks_deno_asymp}. The condition for the leading term in $\tilde{k}$ to survive is
\begin{equation}
    \xi=\left\{\begin{array}{ll}
        \displaystyle 1/(\mu-1)\;, &2<\mu\leq 3 \;, \\
        \displaystyle 1/2\;, &\mu>3 \;. \\
    \end{array}\right.
\end{equation}
Then the limit~\eqref{eqn_def_asym} gives the following.
\begin{equation}
g(\tilde{k},\tilde{s})=\left\{\begin{array}{ll}
    \displaystyle \frac{1}{\tilde{s}+K_\mu\tilde{k}^{\mu-1}}\;, &2<\mu<3 \;, \\[12pt] 
    \displaystyle \frac{1}{\tilde{s}+\tilde{k}^2/[d(\mu-3)]}\;, &\mu>3 \;, \\[12pt] 
\end{array}\right.\label{eqn_asym_alp_dimless_tauT0}
\end{equation}
where $K_\mu:=-(\mu-1)(\mu-2)A_\mu$. We note that the scaling function $g(\tilde{k},\tilde{s})$ takes a form similar to that of the propagator of 1D L\'evy walk particles~\cite{Zaburdaev:2015}.

The asymptotic function $g(k,\tau)$ can now be calculated by performing the inverse Laplace transform from $\tilde{s}$ to $\tilde{t}$ and recovering spatial-temporal units,
\begin{equation}
    \log g(k,\tau)=\left\{\begin{array}{ll}
        \displaystyle -K_\mu \frac{(v_0\tau_0)^{\mu-1}}{\tau_0}k^{\mu-1}\tau\;, &2<\mu<3 \;, \\[12pt] 
        \displaystyle \left(-\frac{1}{4}-\frac{\gamma}{2}+\log 2\right)v_0^2\tau_0k^2\tau\;, &\mu=3\;, d=2 \;, \\[12pt] 
        \displaystyle \left(\frac{1}{9}-\frac{\gamma}{3}\right)v_0^2\tau_0k^2\tau\;, &\mu=3\;, d=3 \;, \\[12pt] 
        \displaystyle -\frac{v_0^2\tau_0}{d(\mu-3)}k^2\tau\;, &\mu>3 \;, \\[12pt] 
    \end{array}\right.  \label{eqn_asym_alp}
\end{equation}
where $\gamma\simeq 0.5772$ is the Euler's constant.

We test the asymptotic form~\eqref{eqn_asym_alp} in particle simulations in Fig.~\ref{fig:ALP_asymptotic}e. The simulation method is detailed in Sec.~\ref{sec_sim_method}. We find that at small enough $k$, the ISFs measured from particle simulations using Eq.~\eqref{eqn_isf_def} resemble an exponential function $\exp(-\lambda(k)\tau)$. Fitting $\lambda(k)$ of the ISFs with varying $k$, we find that it follows a scaling $k^{\mu-1}$ over a wide range of $k$. However, we note that obtaining the asymptotic function is based on a special rescaling~\eqref{eqn_asym_scale} on both temporal and spatial coordinates, which destroys the information on the mean squared displacement (MSD) of particles if $\xi\neq 1/2$ for $2<\mu<3$. One can refer to Ref.~\cite{Schmiedeberg_2009} for a more detailed discussion on this issue. Thus, the asymptotic function~\eqref{eqn_asym_alp} fails in the vicinity of $k=0$ if $2<\mu<3$ and cannot predict the MSD of particles, which we will calculate in Sec.~\ref{sec_msd}. $K_\mu$ in Eq.~\eqref{eqn_asym_alp} is \textit{not} the generalized diffusion constant of ALPs. 

\subsection{Differences between ALPs and RTPs reveal in large length scales}\label{sec_compare}

We first note that on length scales smaller than $\ell_p$, the reorientation of the particles does not take effect, and the ISFs of both ALPs and RTPs converge to that of a straight swimmer, which equals the swimming propagator $\mathbb{P}_R$. With $D=0$, $\sigma_v=0$, and $\tau_T=0$, $\mathbb{P}_R=J_0(kv_0\tau)$ in a 2D space (Fig.~\ref{fig:ALP_asymptotic}a-d), which is an oscillating function in $\tau$. With decreasing $k$, $f(k,\tau)$ transitions from an oscillatory function $\mathbb{P}_R$ to an exponentially decaying function $g(k,\tau)$. As $\mu$ approaching 2 in ALPs, the transition occurs on a larger length scale (Fig.~\ref{fig:ALP_asymptotic}a-d). The qualitative difference in ISFs of ALPs and RTPs is exhibited by the way the oscillation in $f(k,\tau)$ decays with an increasing length scale. Thus, the asymptotic behavior of the ISF of ALPs reveals qualitative differences from that of RTPs. 

To demonstrate the difference between ALPs and RTPs, let us review the asymptotic behavior of RTP ISF~\cite{Martens:2012}. With the same simplification $D=0$, $\sigma_v=0$, and $\tau_T=0$, and dimensionless variables $\hat{s}:=s\tau_R$ and $\hat{k}:=kv_0\tau_R$, the ISF of RTPs in the Fourier-Laplace domain is $\hat{f}_{\rm RTP}(\hat{k},\hat{s}):=f_{\rm RTP}(k,s)/\tau_R$ and~\cite{Martens:2012,zhao2024quantitative}
\begin{align}
    \hat{f}^{\rm 2D}_{\rm RTP}(\hat{k},\hat{s})=&\frac{1}{\sqrt{(\hat{s}+1)^2+\hat{k}^2}-1}\;, \label{eqn_fks_rtp}\\
    \hat{f}^{\rm 3D}_{\rm RTP}(\hat{k},\hat{s})=&\frac{\arctan(\hat{k}/(\hat{s}+1))}{\hat{k}-\arctan(\hat{k}/(\hat{s}+1))}\;.
\end{align}
With the same scaling~\eqref{eqn_asym_scale}, the asymptotic function $g(k,\tau)$ of $f_{\rm RTP}(k,s)$ exists in the limit $b\to0$ if $\xi=1/2$,
\begin{equation}
    g_{\rm RTP}(k,\tau)=\exp(-v_0^2\tau_R k^2\tau/d)\;,   \label{eqn_asym_rtp}
\end{equation}
which takes the form of an ISF of Brownian particles with a $k^2\tau$ scaling and an effective diffusivity $D^{\rm RTP}_{\rm eff}=v_0^2\tau_R/d$. Fig.~\ref{fig:ALP_asymptotic}d shows that $f_{\rm RTP}(k,\tau)$ quickly converges to $g_{\rm RTP}(k,\tau)$ on length scales longer than $10^1\ell_p$, where the persistence length $\ell_p=v_0\tau_R$ for both ALPs and RTPs.

ALPs with $2<\mu<3$ exhibit an asymptotic behavior distinct from that of RTPs. Despite the scaling $k^{\mu-1}\tau$ instead of the scaling $k^2\tau$ in $g(k,\tau)$, the ISF of ALPs converges to an exponential function on a length scale much longer than that of RTPs (Fig.~\ref{fig:ALP_asymptotic}e). From Fig.~\ref{fig:ALP_asymptotic}a-b, the oscillation in $f(k,\tau)$ persists to a length scale of the order $10^2\ell_p$, one order of magnitude longer than that of RTPs. We note that the oscillations are stronger when $\mu$ approaches 2. But on length scales of the order $\ell_p\sim10\ell_p$, the ISFs of ALPs are qualitatively different from those of RTPs.

If $\mu>3$, ALPs are like diffusive Brownian particles on large scales. Compared to the ISF of Brownian particles, $\exp(-Dk^2\tau)$, it scales as $k^2\tau$ with an effective diffusion constant
\begin{equation}
D^{\rm ALP}_{\rm eff}=\frac{v_0^2\tau_0}{d(\mu-3)}=\frac{v_0^2\tau_R(\mu-2)}{d(\mu-3)}\;,\ \ \ \mbox{for}\ \mu>3\;.\label{eqn_deff_alp_from_gks}
\end{equation}
Thus, ALPs with $\mu>3$ share the asymptotic behavior similar to that of RTPs. However, compared to RTPs with the same mean run time $\tau_R$, we first note that $D^{\rm ALP}_{\rm eff}>D^{\rm RTP}_{\rm eff}$, indicating that ALPs can achieve faster diffusion with the same particle speed and persistence length. Although the ISFs of ALPs with $\mu>3$ will finally converge to $g(k,\tau)$ in Eq.~\eqref{eqn_asym_alp} (Fig.~\ref{fig:ALP_asymptotic}e) with $k\to 0$, it occurs on longer length scales than RTPs (Fig.~\ref{fig:ALP_asymptotic}c-d). At length scales ranging from approximately $\ell_p$ to $10\ell_p$, the oscillation in ISFs of ALPs is more profound than that of RTPs. 

In summary, ISFs of ALPs and RTPs differ in asymptotic behavior on large scales. In particular, on the length scales of the order $10\ell_p$, which is measurable in experiments, ALPs and RTPs already show a significant qualitative difference. In Sec.~\ref{sec_sigmav} we show that the difference remains significant if the speed variability is moderate. Thus, an analysis of ISF on various length scales ranging from the order of $\ell_p$ to $10\ell_p$ is expected to detect ALPs from RTPs.

\subsection{Mean-squared displacement of ALPs}\label{sec_msd}

Eq.~\eqref{eqn_fks_alp_simplest} allows us to calculate the exact mean squared displacement of the simplest ALPs. Assuming that the time $t=0$ is arbitrary in a steady trajectory of a particle, the Laplace transform of the MSD of any spatial coordinate is related to the ISF via
\begin{equation}
    \langle\Delta x^2(s)\rangle=-\left.\pddiff{f(k,s)}{k}\right|_{k=0}\;.
\end{equation}
Substituting Eq.~\eqref{eqn_fks_alp_simplest}, direct calculation gives
\begin{equation}
    \frac{d}{2}\cdot\frac{\langle\Delta x^2(s)\rangle}{v_0^2\tau_0^3}=\frac{s\tau_0+2-\mu+(\mu-2)(\mu-1)W(\tilde{s},0;\mu)}{(s\tau_0)^4}\;.\label{eqn_msd_lap}
\end{equation}
Its inverse Laplace transform can be calculated as
\begin{equation}
    \frac{d}{2}\cdot\langle\Delta x^2(t)\rangle=\frac{(1+t/\tau_0)^{4-\mu}-1}{(\mu-3)(\mu-4)}\left(v_0\tau_0\right)^2+\frac{v_0^2\tau_0t}{\mu-3}\;.\label{eqn_msd_alp_simplest}
\end{equation}

Eq.~\eqref{eqn_msd_alp_simplest} is exact at any $t$ for $\mu>2$ and captures the crossover between the short-time and long-time scaling of ALPs. For short times $t$, we have the expansion for $\mu>2$ that
\begin{equation}
    (1+t/\tau_0)^{4-\mu}-1\sim \frac{4-\mu}{\tau_0}t+\frac{(4-\mu)(3-\mu)}{2\tau_0^2}t^2\;,
\end{equation}
so that in the limit $t\to0$, the particle is ballistic and
\begin{equation}
    d\langle\Delta x^2(t)\rangle\underset{t\to 0}{\sim} (v_0t)^2\;.
\end{equation}
At large times $t$, $(1+t/\tau_0)^{4-\mu}-1\sim (t/\tau_0)^{4-\mu}$, and the scaling is dominated by the larger exponent between $4-\mu$ and 1. Thus, for $t\to\infty$ we have
\begin{equation}
    \frac{d}{2}\cdot\frac{\langle\Delta x^2(t)\rangle}{(v_0\tau_0)^2}\underset{t\to \infty}{\sim} \left\{\begin{array}{ll}
        \displaystyle\frac{1}{(\mu-3)(\mu-4)}\left(\frac{t}{\tau_0}\right)^{4-\mu}\;, & 2<\mu<3 \\[12pt] 
        \displaystyle\frac{t}{\tau_0}\log\left(1+\frac{t}{\tau_0}\right)\;, & \mu=3 \\[12pt] 
        \displaystyle\frac{1}{\mu-3}\left(\frac{t}{\tau_0}\right)\;. & \mu>3 \\
    \end{array}\right.\label{eqn_alp_msd}
\end{equation}

\begin{figure}
    \centering
    \begin{tikzpicture}
    \def\hcolx{4.25cm}
    \def\hfigy{3.175cm}
        \begin{scope}
            \node at (0,0) {\includegraphics{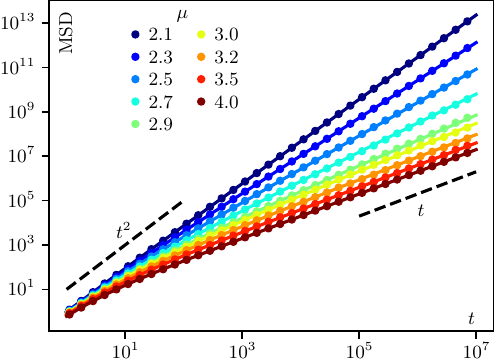}};        
        \end{scope}
    \end{tikzpicture}
    \caption{The mean-squared displacement (MSD) $d\langle\Delta x^2(t)\rangle$ of active L\'evy particles (ALPs) with varying exponent $\mu$ in two dimensional space. Dots represents MSD measured from particle simulations, and solid lines shows the theoretical prediction~\eqref{eqn_alp_msd}. The theoretical prediction~\eqref{eqn_alp_msd} is exact for all time $t$ and exponent $\mu>2$.}
    \label{fig:ALP_MSD}
\end{figure}

For $\mu>3$, the MSD is diffusive, with effective diffusivity $D^{\rm ALP}_{\rm eff}=v_0^2\tau_0/[d(\mu-3)]$, consistent with Eq.~\eqref{eqn_deff_alp_from_gks}. For $2<\mu<3$, the ALPs are superdiffusive and the MSD scales as $t^{4-\mu}$, which is consistent with the results in the literature~\cite{Zaburdaev:2015}. The generalized diffusion constant $\mathcal{D}_\mu$, defined as $\langle\Delta x^2(t)\rangle=2\mathcal{D}_\mu t^{4-\mu}$, is
\begin{equation}
    \mathcal{D}_\mu=\frac{v_0^2\tau_0^{\mu-2}}{d(\mu-3)(\mu-4)}\;.
\end{equation}
We note that unlike RTPs and ALPs with $\mu>3$, $\mathcal{D}_\mu$ is different from the factor in the asymptotic function~\eqref{eqn_asym_alp}. The difference stems from the non-analytic nature of $f(k,s)$ at $(k,s)=(0,0)$ when $2<\mu<3$. Eq.~\eqref{eqn_alp_msd} agrees exactly with the particle simulations in all time $t$ and $\mu>2$ (Fig.~\ref{fig:ALP_MSD}).

\subsection{Active L\'evy particles with finite tumble time}\label{sec_ALP_ft}

We now examine how a finite tumble time will affect the asymptotic behavior of the ISF for ALPs. We keep $D=0$ and $\sigma_v=0$ for simplicity. 

Firstly, assuming the tumble time follows an exponential distribution given in Eq.~\eqref{eqn_alp_varphi}, the Laplace transform of ISF, expressed in dimensionless variables $\tilde{k}=kv_0\tau_0$ and $\tilde{s}=s\tau_0$, is solved as
\begin{equation}
    \frac{f(\tilde{k},\tilde{s})}{p_R\tau_0}=\frac{(\mu-2)[\tau_T/\tau_0+W(\tilde{s},\tilde{k};\mu-1)]^2}{\tau_Ts/\tau_0+1-(\mu-1)W(\tilde{s},\tilde{k};\mu)}+W(\tilde{s},\tilde{k};\mu-2)\;,\label{eqn_isf_ALP_ft}
\end{equation}
where we recall that $p_R=\tau_R/(\tau_R+\tau_T)$ is the fraction of time spent in the runs. Note that setting $\tau_T=0$ recovers Eq.~\eqref{eqn_fks_alp_simplest}. Applying the asymptotic analysis detailed in Sec.~\ref{sec_asymp}, we obtain the dimensionless asymptotic function as
\begin{equation}
g(\tilde{k},\tilde{s})=\left\{\begin{array}{ll}
    \displaystyle \frac{1}{\tilde{s}+p_RK_\mu\tilde{k}^{\mu-1}}\;, &2<\mu<3 \;, \\[12pt] 
    \displaystyle \frac{1}{\tilde{s}+p_R\tilde{k}^2/[d(\mu-3)]}\;, &\mu>3 \;. \\[12pt] 
\end{array}\right.\label{eqn_asymp_alp_ft}
\end{equation}
Comparing Eq.~\eqref{eqn_asymp_alp_ft} to Eq.~\eqref{eqn_asym_alp_dimless_tauT0}, we note that an exponentially distributed finite tumble time does not alter the asymptotic behavior of ISF. The (generalized) diffusion constant is simply rescaled by the factor $p_R$.

\begin{figure}
    \centering
    \begin{tikzpicture}
    \def\hcolx{4.12cm}
    \def\hfigy{3.25cm}

        \begin{scope}[xshift=-0.22*\hcolx,yshift=-1.15*\hfigy]
            \node at (-1.5,1.8) {(a)};
            \node at (0,0) {\includegraphics{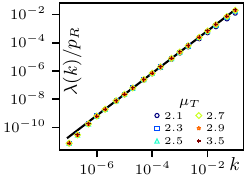}};        
        \end{scope}
        
        \begin{scope}[xshift=0.85*\hcolx,yshift=-1.15*\hfigy]
            \node at (-1.5,1.8) {(b)};
            \node at (0,0) {\includegraphics{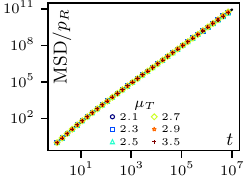}};        
        \end{scope}

    \end{tikzpicture}
    \caption{Asymptotic behavior of intermediate scattering functions (ISFs) of particles with run and tumble times both following Lomax distribution, referred to as Lomax-Lomax (LL) particles, and their mean-squared displacement (MSD). \textbf{(a)} The decaying rate $\lambda(k)$ of ISFs of LL particles as a function of $k$ rescaled by the fraction of time spent in runs $p_R$, where we fit the ISFs of ALPs by an exponential function $\exp(-\lambda(k)\tau)$. The symbols represent $\lambda(k)$ measured from numerical ISFs calculated using Eq.~\eqref{eqn_isf_def}. The black line shows the asymptotic decaying rate in Eq.~\eqref{eqn_asymp_alp_ft}. \textbf{(b)} The MSD of the LL particles rescaled by $p_R$. The black line represent Eq.~\eqref{eqn_alp_msd_ft}. Note that the finite tumble time only rescales time by $p_R$ in the ISF and MSD, regardless of $\mu_T$. Color encodes $\mu_T$. Parameters in panels (a) and (b): $v_0=1$, $\tau_0=\tau_{T0}=1$, $\mu=2.5$, $d=2$.}
    \label{fig:isf_ft}
\end{figure}

Next, we consider a tumble time distribution with a heavy tail, modeled by a Lomax distribution
\begin{equation}
    \varphi_T^{\rm LL}(\tau)=\frac{\mu_T-1}{\tau_{T0}}\left(1+\frac{\tau}{\tau_{T0}}\right)^{-\mu_T}\;.
\end{equation}
Now, both run and tumble times follow Lomax distributions with different exponents and cut-off times, which we refer to as Lomax-Lomax (LL) particles. The renewal theory remains applicable for $\mu>2$ and $\mu_T>2$, where the mean run and tumble times are finite. In this regime, the ISF is
\begin{widetext}
\begin{align}
    &f(\tilde{k},\tilde{s})=p_R\tau_0W(\tilde{s},\tilde{k};\mu-2)+p_T\tau_{T0}W_0(\beta\tilde{s};\mu_T-2)+ \nonumber \\
    &\frac{\tau_0^2(\mu_T-1)W(\tilde{s},\tilde{k};\mu-1)^2W_0(\beta\tilde{s};\mu_T)+\tau_{T0}^2(\mu-1)W_0(\beta\tilde{s};\mu_T-1)^2W(\tilde{s},\tilde{k};\mu)+2\tau_0\tau_{T0}W(\tilde{s},\tilde{k};\mu-1)W_0(\beta\tilde{s};\mu_T-1)}{[1-(\mu-1)(\mu_T-1)W(\tilde{s},\tilde{k};\mu)W_0(\beta\tilde{s};\mu_T)](\tau_R+\tau_T)}\;,\label{eqn_isf_LL}
\end{align}
\end{widetext}
where $W_0(x;\nu)=W(x,0;\nu)$, $\beta=\tau_{T0}/\tau_0$ is the ratio between characteristic tumble and run times, and $p_T=1-p_R$ is the fraction of time spent in tumbles. Although Eq.~\eqref{eqn_isf_LL} contains more terms than Eq.~\eqref{eqn_isf_ALP_ft} and~\eqref{eqn_fks_alp_simplest}, the same asymptotic analysis yields the identical results as Eq.~\eqref{eqn_asymp_alp_ft}. This is verified in Fig.~\ref{fig:isf_ft}a using particle simulations. Thus, the specific form of the tumble time distribution does not qualitatively modify the asymptotic function of the ISF, provided the mean tumble time is finite. This invariance stems from the fact that the leading-order term of $\mathcal{L}[\varphi_T(\tau)\mathbb{P}_T(k,\tau)](s)$ in $k$ is always analytic even if $D$ is non-zero.

Following the same procedure in Sec.~\ref{sec_msd}, the mean-squared displacement of LL particles is calculated as
\begin{equation}
    \frac{d\langle\Delta x^2(t)\rangle}{2p_R}=\frac{(1+t/\tau_0)^{4-\mu}-1}{(\mu-3)(\mu-4)}\left(v_0\tau_0\right)^2+\frac{v_0^2\tau_0t}{\mu-3}\;,\label{eqn_alp_msd_ft}
\end{equation}
which is identical to Eq.~\eqref{eqn_alp_msd} scaled by the fraction of run time $p_R$. Thus, for $\mu_T>2$, the long-time diffusive behavior of an LL particle is determined solely by $\mu$, consistent with Ref.~\cite{datta_random_2024}. This is further confirmed by particle simulations in Fig.~\ref{fig:isf_ft}b.

In summary, a finite tumble time only rescales the effective time by a factor of $p_R$ at large scales, regardless of its specific distribution, as long as its mean is finite. As a result, we do not expect to detect a possible heavy-tailed tumble time distribution from experimental measurements of the ISF. We note that the tumble time distribution only influences the diffusion exponent if the mean tumble time diverges ($\mu_T<2$)~\cite{datta_random_2024}.

\subsection{Large speed variability suppresses differences between ALPs and RTPs}\label{sec_sigmav}

\begin{figure*}
    \centering
    \includegraphics{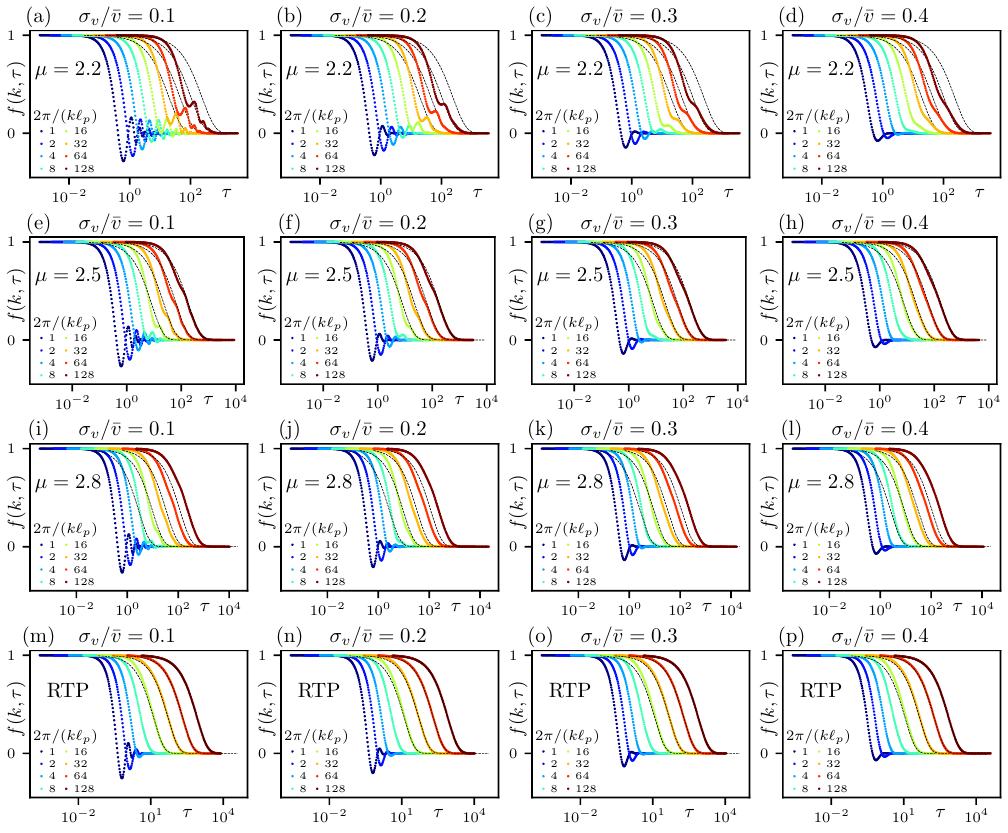}
    \caption{Intermediate scattering functions (ISFs) of 2D \textbf{(a)-(l)} active L\'evy particles (ALPs) and \textbf{(m)-(p)} run-and-tumble particles (RTPs) with varying standard deviation $\sigma_v$ of speed. We fix the run time $\tau_R=1$ in all panels. Rows represent ALPs with exponent $\mu=2.2$ (a)-(d), 2.5 (e)-(h), 2.8 (i)-(l), or RTPs (m)-(p). Columns represent $\sigma_v=0.1$ (a,e,i,m), 0.2 (b,f,j,n), 0.3 (c,g,k,o), and 0.4 (d,h,l,p). Circles represent ISF calculated using Eq.~\eqref{eqn_isf_def} from particle simulations, and black dashed lines show the asymptotic functions for ALPs (Eq.~\eqref{eqn_asym_alp_sigma}) and RTPs ($g(k,\tau)=\exp[-(\bar{v}^2+\sigma_v^2)\tau_Rk^2\tau/2]$) for $2\pi/k=16\ell_p$, $32\ell_p$, $64\ell_p$, $128\ell_p$. We note that with increasing $\sigma_v$, oscillations in ISFs are surpressed, and the difference of ALPs and RTPs becomes less significant. Other parameters: $\bar{v}=1$, $\tau_T=0$, $D=0$.}
    \label{fig_ISF_ALP_sigma}
\end{figure*}

In this section, we investigate the effect of speed variability, characterized by the standard deviation of the run speed $\sigma_v$, on the precision of the DDM analysis. We assume instantaneous tumbles ($\tau_T=0$) and set the the thermal translational diffusion coefficient $D=0$. First, we note that a finite $\sigma_v$ does not qualitatively alter the asymptotic behavior of ALPs. In the expansion of the run propagator $\mathbb{P}_R$ in terms of $k$, the integration operator $\int p(v;\bar{v},\sigma_{v})dv$ modifies the coefficients by replacing the power functions $v^\nu$ with their average $\langle v^\nu\rangle$, taken with respect to the speed distribution $p(v;\bar{v},\sigma_v)$. For the Schultz distribution that we consider,
\begin{equation}
    \langle v^\nu\rangle=\frac{\Gamma(\nu+\bar{v}^2/\sigma_v^2)}{\Gamma(\bar{v}^2/\sigma_v^2)}\left(\frac{\sigma_v^2}{\bar{v}}\right)^\nu\;.
\end{equation}
Setting $\nu=2$ yields $\langle v^2\rangle=\bar{v}^2+\sigma_v^2$~\cite{kurzthaler2024characterization}. Consequently, a finite $\sigma_v$ shifts the (generalized) diffusion coefficient at large scales, and the asymptotic function of the ISF of ALPs in Eq.~\eqref{eqn_asym_alp} becomes
\begin{equation}
    \log g(k,\tau)=\left\{\begin{array}{ll}
        \displaystyle -K_\mu \frac{\langle v^{\mu-1}\rangle\tau_0^{\mu-1}}{\tau_0}k^{\mu-1}\tau\;, &2<\mu<3 \;, \\[12pt] 
        \displaystyle -\frac{(\bar{v}^2+\sigma_v^2)\tau_0}{d(\mu-3)}k^2\tau\;, &\mu>3 \;. \\[12pt] 
    \end{array}\right.  \label{eqn_asym_alp_sigma}
\end{equation}

However, a large $\sigma_v$ suppresses the oscillations in the ISF, making it more challenging to distinguish between ALPs and RTPs. As shown in Fig.~\ref{fig_ISF_ALP_sigma}, which compares the ISFs of ALPs and RTPs for increasing values of $\sigma_v$, the ISFs of ALPs enter an exponential decaying regime at shorter length scales as $\sigma_v$ increases, thereby diminishing the observable differences between the two models. This phenomenon is similar to the findings reported in Ref.~\cite{petrovskii_variation_2011}. Although the qualitative asymptotic behavior of ALPs remains unaffected by a large $\sigma_v$, extracting the scaling exponent in $k$ requires accessing data at length scales of orders of magnitude larger than $10\ell_p$ within the asymptotic scaling. Thus, we expect that DDM analysis can reliably differentiate ALPs in experimental data for $k\lesssim 10\ell_p$ only if the speed variability is sufficiently small ($\sigma_v\lesssim 0.2\bar{v}$). Otherwise, it is necessary to probe length scales over a wider range of orders of magnitude.

\section{Detecting active L\'evy particles using differential dynamic microscopy}\label{sec_validation}

\begin{figure*}
    \centering
    \begin{tikzpicture}
    \def\hcolx{4.12cm}
    \def\hfigy{3.25cm}
        \begin{scope}[xshift=-1*\hcolx]
            \node at (-1.8,2) {(a)};
            \node at (0,0) {\includegraphics{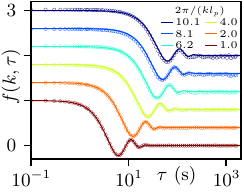}};
            \node[font=\large] at (0.45,2.0) {$\mu=2.05$};
            \node[rotate=90,font=\large] at (-2.5,0.2) {\textbf{ALP Model}};
        \end{scope}
        
        \begin{scope}[xshift=0.05*\hcolx]
            \node at (-1.8,2) {(b)};
            \node at (0,0) {\includegraphics{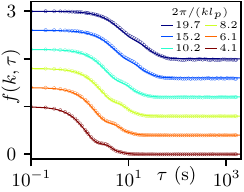}};
            \node[font=\large] at (0.45,2.0) {$\mu=2.6$};
        \end{scope}
        
        \begin{scope}[xshift=1.1*\hcolx]
            \node at (-1.8,2) {(c)};
            \node at (0,0) {\includegraphics{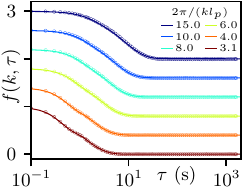}};
            \node[font=\large] at (0.45,2.0) {$\mu=3.5$};
        \end{scope}

        \begin{scope}[xshift=2.15*\hcolx]
            \node at (-1.8,2) {(d)};
            \node at (0,0) {\includegraphics{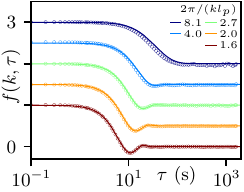}};
            \node[font=\large] at (0.4,2.0) {RTPs};
        \end{scope}

        \begin{scope}[xshift=-1*\hcolx,yshift=-1.2*\hfigy]
            \node at (-1.8,2) {(e)};
            \node at (0,0) {\includegraphics{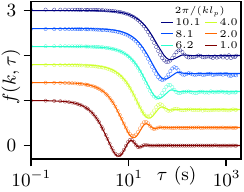}}; 
            \node[rotate=90,font=\large] at (-2.5,0.2) {\textbf{RTP Model}};
        \end{scope}
        
        \begin{scope}[xshift=0.05*\hcolx,yshift=-1.2*\hfigy]
            \node at (-1.8,2) {(f)};
            \node at (0,0) {\includegraphics{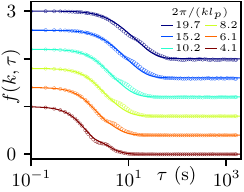}}; 
        \end{scope}
        
        \begin{scope}[xshift=1.1*\hcolx,yshift=-1.2*\hfigy]
            \node at (-1.8,2) {(g)};
            \node at (0,0) {\includegraphics{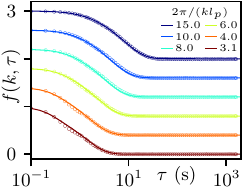}};
        \end{scope}

         \begin{scope}[xshift=2.15*\hcolx,yshift=-1.2*\hfigy]
            \node at (-1.8,2) {(h)};
            \node at (0,0) {\includegraphics{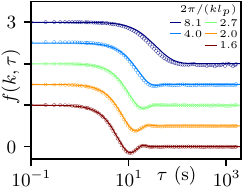}};
        \end{scope}

        \begin{scope}[xshift=-1.01*\hcolx,yshift=-2.4*\hfigy]
            \node at (-1.7,2) {(i)};
            \node at (0,0) {\includegraphics{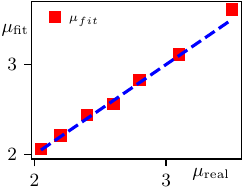}};
        \end{scope}
        
        \begin{scope}[xshift=0.01*\hcolx,yshift=-2.4*\hfigy]
            \node at (-1.7,2) {(j)};
            \node at (0,0) {\includegraphics{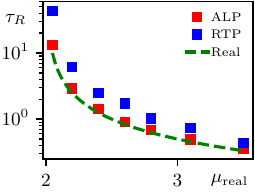}};        
        \end{scope}
        
        \begin{scope}[xshift=1.04*\hcolx,yshift=-2.4*\hfigy]
            \node at (-1.55,2) {(k)};
            \node at (0,0) {\includegraphics{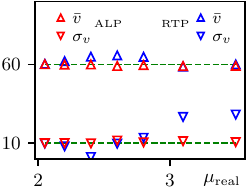}};        
        \end{scope}

        \begin{scope}[xshift=2.1*\hcolx,yshift=-2.4*\hfigy]
            \node at (-1.55,2) {(l)};
            \node at (0,0) {\includegraphics{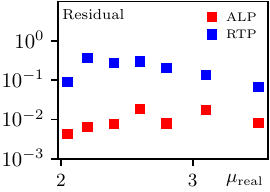}};        
        \end{scope}
        
    \end{tikzpicture}
    \caption{Validation of the fitting protocol using simulations. The ISFs calculated from the synthetic images are shown in \textbf{(a,b,c,e,f,g)} for ALPs and in \textbf{(d,h)} for RTPs. The ISFs are shifted vertically for better visualization. Symbols represent the ISFs measured from simulations, and lines are fits using the ALP model (a-d) and the RTP model (e-g). Colors encode wavenumber $k$. \textbf{(i)} Comparison between the fitted exponent $\mu_{\rm fit}$ and the ground truth $\mu_{\rm real}$. The dashed blue line shows the position for the perfect fit $\mu_{\rm fit}=\mu_{\rm real}$. \textbf{(j)} The estimates of mean run time $\tau_R$ of ALPs from the ALP model and the RTP model. The dashed green line represents the ground truth. \textbf{(k)} The estimates of $\bar{v}$ and $\sigma_v$ of ALPs using both models. The green dashed lines show the ground truth of both parameters. \textbf{(l)} The residue of the fitting of ISFs of ALPs using both models. The residue is defined as $\sum |f_{\rm fit}(k,\tau)-f_{\rm data}(k,\tau)|^2$ where the summation is over all points considered in fitting. The correct model generally fits better the synthetic data and provide estimates close to the ground truth.}
    \label{fig:fit_isf_sim_alp}
\end{figure*}

\subsection{Differential dynamic microscopy}\label{sec_ddm}
Now we provide a protocol for using differential dynamic microscopy to characterize the motion of cells. We take bright-field time-lapse images of the cells swimming in the liquid. Cells can swim in a quasi-2D or 3D environment. Assuming $I(\rr,t)$ is the intensity of pixels at position $\rr$ and time $t$, we calculate
\begin{equation}
    \Delta I(\rr,t)=I(\rr,t)-\langle I(\rr,t)\rangle_t\;,
\end{equation}
where $\langle I(\rr,t)\rangle_t$ is the average intensity of the pixels over time. Then we calculate the Fourier transformation of $\Delta I(\rr,t)$ and average over the direction of the wave vector $\kk$. We assume $\Delta I\propto \Delta\rho$ where $\rho$ is the density of cells. Instead of calculating the differential intensity correlation function as in Ref.~\cite{Wilson:2011,Martinez:2012,zhao2024quantitative}, we directly calculate the intermediate scattering function as the autocorrelation function of $\Delta I(k,t)$,
\begin{equation}
    f(k,\tau)=\frac{\langle\Delta I(k,t)\Delta I^*(k,t+\tau)\rangle_t}{\langle|\Delta I(k,t)|^2\rangle_t}\;.\label{eqn_isf_acorr}
\end{equation}
Note that this approach is equivalent to that adopted in Ref.~\cite{Wilson:2011,Martinez:2012,zhao2024quantitative} but does not require reaching the plateau at $\tau=0$ and $\tau\to\infty$. Then the experimentally measured $f(k,\tau)$ is fitted by its theoretical prediction
\begin{equation}
    f(k,\tau)=\alpha f_{\rm RT}(k,\tau)+(1-\alpha)\exp(-Dk^2\tau)\;,
\end{equation}
where we add a fraction $1-\alpha$ of nonmotile diffusive cells with the same diffusion constant $D$ as motile cells, and $f_{\rm RT}(k,\tau)$ is the ISF of ALPs or RTPs calculated using Eq.~\eqref{eqn_isf_rt} and the renewal theory.

To characterize the motion of cells across scales, we fit simultaneously $f(k,\tau)$ with multiple $k$ values that range across scales. The choices of $k$ should include those on small length scales $\sim \ell_p$, which encode swimming properties, and those on large length scales at least $5\ell_p$, which capture the asymptotic behavior of $f(k,\tau)$ on large scales. We note that the smallest $k$ accessible experimentally is constrained by the image size in quasi-2D systems or the depth of fields in 3D systems~\cite{Martinez:2012}, and the measurement on large length scales approaching $1/8$ of the image size suffers from insufficient statistical accuracy. The largest $k$ is restricted by the pixel size of the images. As visualization of a single cell is not necessary, using a pixel size between the body size of the cell and $\ell_p$ should be sufficient.

\subsection{Validations of the method using simulation-generated data}\label{sec_sim_method}

We first validate our protocol using simulations, where we know the ground truth of the cell motion. To simulate active L\'evy particles, we sequentially sample the time $\tau$ of the next run or tumble event, according to the corresponding probability distributions $\varphi_{R,T}(\tau)$. The particle position is then updated to the next snapshot time or $\tau$. We assume that the particles are initially stationary and distributed uniformly in the simulation space. We note that unlike RTPs where the run and tumble processes are Poissonian, the initial run-time distribution of ALPs is not the Lomax distribution $\varphi_R^{\rm ALP}(\tau)$~\eqref{eqn_alp_varphi} but~\cite{Detcheverry:2015}
\begin{equation}
    \varphi_{R,0}^{\rm ALP}(\tau):=\int_\tau^\infty \frac{\varphi_R^{\rm ALP}(t)}{\tau_R}dt=\frac{\mu-2}{\tau_0(1+\tau/\tau_0)^{\mu-1}}
\end{equation}
Then we generate artificial imaging data using the same method as in Ref.~\cite{zhao2024quantitative}. $f(k,\tau)$ is calculated and fitted using the protocol described in Sec.~\ref{sec_ddm}. 

Specifically, we simulate ALPs in 2D periodic systems with kinetic parameters similar to those measured from tracking Euglena cells~\cite{li2025biased}, $\alpha=1.0$, $\bar{v}=60  {\rm\ \upmu m/s}$, $\sigma_v=10 {\rm\ \upmu m/s}$, $\tau_0=0.5 {\rm\ s}$, $\tau_T=0.5 {\rm\ s}$, $D=0 {\rm\ \upmu m^2/s}$. We vary $\mu$ from $2.05$ to $3.5$ with fixed image size of $L=4096$ pixels and pixel size $\delta x=9.6 \upmu$m. The image size and resolution are accessible in experiments. Note that the samples have varying $\ell_p$ while $L$ is fixed. As $\ell_p$ is unknown \textit{a priori} in real experiments, we do not adjust the image size, in order to test the robustness of our protocol. The size of the simulation box $\tilde{L}=5000$ pixel is larger than the image size to avoid possible correlations due to the periodic boundaries. The ISFs are then calculated using Eq.~\eqref{eqn_isf_acorr}.

We then fit the kinetic parameters, $(\bar{v}, \sigma_v,\tau_0,\mu,\tau_T)$, from the ISFs calculated from the synthetic images using the Levenberg-Marquardt algorithm~\cite{levenberg1944method}. We fix $\alpha=1$ and $D=0$ because the \textit{E. gracilis} cells are large, so their thermal diffusion is negligible. We fit simultaneously 10 ISFs with different $k$ values, chosen by Latin hypercube sampling in the range $k\in(0.0013,0.1014){\rm\ \upmu m^{-1}}$. The fits with the smallest residue $\sum_{k,\tau} |f_{\rm fit}(k,\tau)-f_{\rm data}(k,\tau)|^2$ are chosen as the parameter estimations. The fitting results are then compared with the ground truth. The ISFs and the fitting results are shown in Fig.~\ref{fig:fit_isf_sim_alp}. We note that the ALP model fits well the synthetic imaging data of the ALP (Fig.~\ref{fig:fit_isf_sim_alp}a-c), with reliable extraction of the exponent $\mu$ and $\tau_R$ (Fig.~\ref{fig:fit_isf_sim_alp}i-j). The fitting using the RTP model is generally worse than using the correct model (Fig.~\ref{fig:fit_isf_sim_alp}e-g,l), estimating a larger $\tau_R$ than the ground truth (Fig.~\ref{fig:fit_isf_sim_alp}j). Notably, the RTP model generally overlooks the persistent oscillations in $f(k,\tau)$ of ALPs on large length scales. Both models give similar estimations of other parameters, for example $\bar{v}$ and $\sigma_v$, while the error of the estimations from the wrong model is larger (Fig.~\ref{fig:fit_isf_sim_alp}k). 

In contrast, it is possible that the ALP model fits the ISF of RTP better than the correct RTP model, since the ALP model has one more parameter than the RTP model, and the difference between the two models vanishes at $\mu\to\infty$ (Fig.~\ref{fig:fit_isf_sim_alp}d,h). However, the fitted $\mu=650\gg 3$ suggests an overfitting, indicating that the ALP model is not a minimal description of the data.

In particular, the protocol reliably extracts $\mu$ that range from 2 to more than 3 and $\tau_R$ that span over 1 decade, when the speed variability is moderate with $\sigma_v/\bar{v}=0.17$. The protocol is robust when fitting the length scales over $1\ell_p\sim 30\ell_p$. The difference in the goodness of fitting between the ALP and RTP models decreases when $\mu>3$, as the difference between ALPs and RTPs diminishes with $\mu\to\infty$.

\section{Detecting L\'evy walk in experimental data}\label{sec_exp}

In this section, we apply our protocol to the experimental data of a flagellated bacterial \textit{E. coli} and an algae \textit{E. gracilis}. For \textit{E. coli}, we use the published data in Ref.~\cite{kurzthaler2024characterization}.

\subsection{\textit{E. coli}}

\begin{figure}
    \centering
    \begin{tikzpicture}
    \def\hcolx{4.12cm}
    \def\hfigy{3.25cm}
        \begin{scope}[xshift=-0.22*\hcolx]
            \node at (-1.5,1.8) {(a)};
            \node at (0,0) {\includegraphics{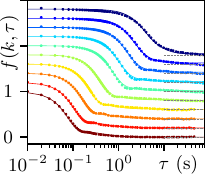}};    
            \node at (0.2,1.8) {ALP model};
        \end{scope}
        
        \begin{scope}[xshift=0.35*\hcolx,yshift=0.4cm]
            \node at (0,0) {\includegraphics{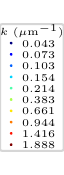}}; 
        \end{scope}
        
        \begin{scope}[xshift=0.94*\hcolx]
            \node at (-1.5,1.8) {(b)};
            \node at (0,0) {\includegraphics{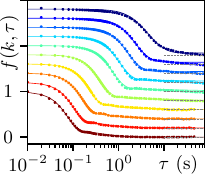}};     
            \node at (0.2,1.8) {RTP model};         
        \end{scope}
        
        \begin{scope}[xshift=-0.22*\hcolx,yshift=-1.15*\hfigy]
            \node at (-1.5,1.8) {(c)};
            \node at (0,0) {\includegraphics{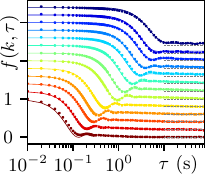}};    
            \node at (0.2,1.8) {ALP model};
        \end{scope}
        
        \begin{scope}[xshift=0.35*\hcolx,yshift=0.12cm-1.15*\hfigy]
            \node at (0,0) {\includegraphics{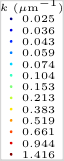}}; 
        \end{scope}
        
        \begin{scope}[xshift=0.94*\hcolx,yshift=-1.15*\hfigy]
            \node at (-1.5,1.8) {(d)};
            \node at (0,0) {\includegraphics{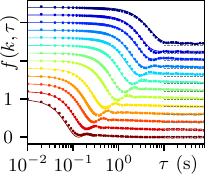}};     
            \node at (0.2,1.8) {RTP model};         
        \end{scope}

        \begin{scope}[xshift=-0.22*\hcolx,yshift=-2.3*\hfigy]
            \node at (-1.5,1.8) {(e)};
            \node at (0,0) {\includegraphics{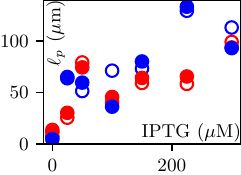}};        
        \end{scope}
        
        \begin{scope}[xshift=0.89*\hcolx,yshift=-2.3*\hfigy]
            \node at (-1.5,1.8) {(f)};
            \node at (0,0) {\includegraphics{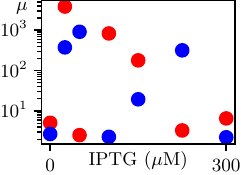}};        
        \end{scope}

        \begin{scope}[xshift=0.3*\hcolx,yshift=-3.4*\hfigy]
            \node at (-3.65,1.8) {(g)};
            \node at (0,0) {\includegraphics{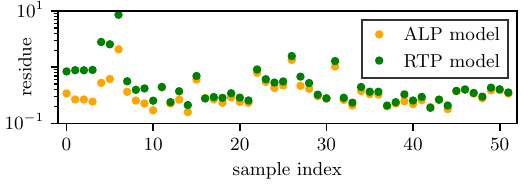}};
        \end{scope}
    \end{tikzpicture}
    \caption{Fitting results of the ISFs of \textit{E. coli} cells using the active L\'evy particle model. \textbf{(a-b)} Comparison of the ISFs of wild-type \textit{E. coli} fitted by (a) the ALP model and (b) the RTP model, correspondingly. The fitting results from the RTP model is taken from Ref.~\cite{kurzthaler2024characterization}. \textbf{(c-d)} Comparison of the ISFs of \textit{E. coli} strain NZ1 with IPTG concentration 150 $\upmu$M fitted by (c) the ALP model and (d) the RTP model, correspondingly. In panels (a)-(d), the ISFs are shifted vertically and gray dashed lines correspond to $f=0$. Symbols represent experimental data and lines are fits to the theory. Colors encode wavenumber $k$. \textbf{(e)} Comparison of fitted persistence length $\ell_p$ with respect to different IPTG concentration from the ALP model (solid symbols) and the RTP model (open symbols). \textbf{(f)} The fitted exponent $\mu$ in the ALP model as a function of IPTG concentration. The ISF data and fits of the RTP model are taken from Ref.~\cite{kurzthaler2024characterization}. Red and blue symbols in panels c, d correspond to two biological replicates. \textbf{(g)} The residue of fitting of all samples. The residue is defined as $\sum |f_{\rm fit}(k,\tau)-f_{\rm data}(k,\tau)|^2$ where the summation is over all points considered in fitting.}
    \label{fig:ecoli}
\end{figure}

Ref.~\cite{kurzthaler2024characterization} has reported that the ISFs of \textit{E. coli} are well fitted by the RTP model. Now we test if the ALP model fits the same data. We first fit the ISFs of wild-type (WT) AB1157 \textit{E. coli} using the ALP model. We notice that the ALP model can reach a similar goodness of fitting as the RTP model (Fig.~\ref{fig:ecoli}a,b). The fitted parameters in the ALP model are $\alpha=0.975\pm0.005$, $\bar{v}=15.7\pm0.4{\ \rm \upmu m\ s^{-1}}$, $\sigma_v=5.6\pm0.4{\ \rm \upmu m\ s^{-1}}$, $\tau_0=1.60\pm0.50{\ \rm s}$, $\mu=2.65\pm0.32$, $\tau_T=0.4\pm0.1{\ \rm s}$, $D=0.20\pm0.05{\ \rm \upmu m^2\ s^{-1}}$, leading to $\tau_R=2.45\pm 1.44{\ \rm s}$. Errors are obtained by a jackknife resampling method~\cite{Efron:1981}. Compared to the fitting result of the RTP model as published in Ref.~\cite{kurzthaler2024characterization}, $\alpha=0.96\pm0.001$, $\bar{v}=16\pm0.1{\ \rm \upmu m\ s^{-1}}$, $\sigma_v=5.78\pm 0.13{\ \rm \upmu m\ s^{-1}}$, $\tau_R=2.39\pm0.11{\ \rm s}$, $\tau_T=0.38\pm0.02{\ \rm s}$, and $D=0.24\pm0.01{\ \rm \upmu m^2\ s^{-1}}$, the two models provide similar mean estimations, but the ALP model shows a large estimation error in jackknife resampling. The large errors indicate that the goodness of fitting is insensitive to the particular values of $\tau_0$ and $\mu$, which suggests an overfitting of the ALP model. Thus, the estimated $\mu<3$ should \textit{not} be considered as an evidence for the L\'evy walk of wild-type \ecoli. Because the cells have a large speed variability $\sigma_v/\bar{v}=0.36$, as seen from the lack of oscillations in ISFs on short length scales, the DDM analysis cannot reliably distinguish ALPs from the experimental data of $2\pi/k\lesssim 4\ell_p$, which is much shorter than the asymptotic regime (see Sec.~\ref{sec_sigmav}).

We then fit the ISFs of the engineered NZ1 strain with the ALP model, where cell motility is controlled by an external inducer Isopropyl $\beta$-d1-thiogalactopyranoside (IPTG). A sample of ISF and the estimated parameters are shown in Fig.~\ref{fig:ecoli}c-g. NZ1 strain has a longer persistence length than the WT strain, and the oscillations in the ISF are more significant. The ALP model is able to fit the measured ISFs of \textit{E. coli} (Fig.~\ref{fig:ecoli}a,b), giving estimates of swimming speed and persistence length consistent with those from the RTP model reported in Ref.~\cite{kurzthaler2024characterization} (Fig.~\ref{fig:ecoli}c). The goodness of fitting slightly improves with the ALP model (Fig.~\ref{fig:ecoli}e). However, the estimated exponents $\mu$ are unstable and generally of orders of magnitude larger than 3 (Fig.~\ref{fig:ecoli}d), undermining the possibility of describing swimming \textit{E. coli} as active L\'evy particles. This is also seen in the asymptotic behavior of the \textit{E. coli} ISFs (Fig.~\ref{fig:ecoli}a,b). The oscillation in ISFs of large $k$ decays quickly as $k$ decreases, which is only possible for RTPs or ALPs with large enough $\mu$. 

These results show that the ISF of RTPs can be well approximated by that of ALPs with extremely large $\mu$, suggesting redundant parameters and overfitting. Thus, the motion of \textit{E. coli} in 3D bulk liquid is better described as RTPs.

\subsection{\textit{E. gracilis}}
\begin{figure}
    \centering
    \begin{tikzpicture}
    \def\hcolx{4.12cm}
    \def\hfigy{3.25cm}
        \begin{scope}[xshift=-0.2*\hcolx]
            \node at (-1.8,2) {(a)};
            \node at (0,0) {\includegraphics{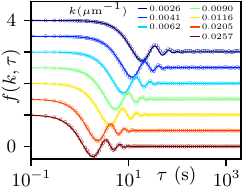}};       
            \node at (0.2,2) {ALP model};
            \node at (-1.2,-0.9) {$0$};
        \end{scope}
        
        \begin{scope}[xshift=0.88*\hcolx]
            \node at (-1.8,2) {(b)};
            \node at (0,0) {\includegraphics{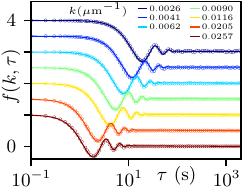}};       
            \node at (0.2,2) {RTP model};           
        \end{scope}

        \begin{scope}[xshift=-0.2*\hcolx,yshift=-1.2*\hfigy]
            \node at (-1.8,2) {(c)};
            \node at (0,0) {\includegraphics{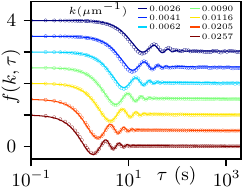}};     
            \node at (-1.2,-0.9) {$0.3$};   
        \end{scope}
        
        \begin{scope}[xshift=0.88*\hcolx,yshift=-1.2*\hfigy]
            \node at (-1.8,2) {(d)};
            \node at (0,0) {\includegraphics{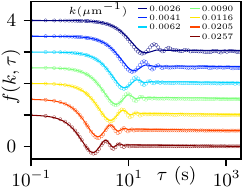}};        
        \end{scope}

        \begin{scope}[xshift=-0.2*\hcolx,yshift=-2.4*\hfigy]
            \node at (-1.8,2) {(e)};
            \node at (0,0) {\includegraphics{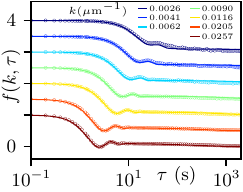}};    
            \node at (-1.2,-0.9) {$0.6$};    
        \end{scope}
        
        \begin{scope}[xshift=0.88*\hcolx,yshift=-2.4*\hfigy]
            \node at (-1.8,2) {(f)};
            \node at (0,0) {\includegraphics{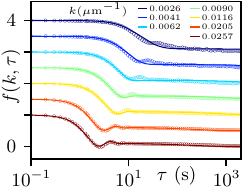}};        
        \end{scope}
    \end{tikzpicture}
    \caption{The ISFs of \textit{E. gracilis} with three light intensities: \textbf{(a,b)} 0 W/m$^2$, \textbf{(c,d)} 0.3 W/m$^2$, \textbf{(e,f)} 0.6 W/m$^2$, labeled in the figures. The ISFs are shifted vertically for better visualization. Symbols represent experimental data, and lines shows fits from the ALP model (a,c,e) and the RTP model (b,d,f). Color encodes wavenumber $k$.}
    \label{fig:Fit_isf_euglena}
\end{figure}

The \textit{E. gracilis} cells used in our experiments are cultured in modified Cramer-Myers medium under continuous illumination of 7 \si{W/m^2} at 25 \si{\celsius}~\cite{li2025biased}. The samples are introduced into a rectangular chamber of size 40  \si{mm} $\times$40 \si{mm} $\times$ 150\si{\micro m}, constructed from two glass slides joined with double-sided waterproof tape.

To ensure uniform illumination, the setup is placed in a dark room kept at 25 \si{\celsius}. A Texas Instruments DLP4710G2 projector, combined with an 85-mm lens, delivers blue-light illumination (470 \si{nm}). The intensity of the light is adjusted by modulating the current of the projector's LEDs or varying the intensity of the blue channel in the input image stream. Measurements of the actual light intensity in the sample plane are obtained using a digital optical power meter (DHC GCI-080102). Infrared illumination (850 \si{nm}) is used to provide homogeneous background lighting without affecting the phototaxis behavior of cells. 

Cell trajectories are recorded using a Teledyne DALSA Genie Nano-CL M4090 CMOS area scan camera (16 MP) equipped with an object space telecentric lens (ES model ESCM045-191D175) at 0.45$\times$ magnification, operating at 10 frames per second. This imaging system provides a full view of the entire sample chamber, while maintaining sufficient resolution to capture intensity fluctuations arising from the motion of individual cells. Each captured image is Fourier-transformed and averaged over directions of $\kk$ to calculate the ISF.

Then we fit the ISFs of \textit{E. gracilis} under different light intensities $I$ using both the ALP model and the RTP model. Fig.~\ref{fig:Fit_isf_euglena} shows the comparison between the fitting results of the ALP model and the RTP model. For cells in a light intensity $I>0$, the ALP model fits the experimental data better than the RTP model. The ISFs of \textit{E. gracilis} exhibit a stronger oscillation at small $k$, which is difficult to be captured by the RTP model. We note that both the ALP model and the RTP model fit the data with $I=0$ W/m$^2$. Because the cells with $I=0$ W/m$^2$ are estimated to have a large persistence length $\ell_p$ comparable to the image size (Tab.~\ref{tab:euglena_results}), the measured ISF does not reach a length scale of several $\ell_p$, so that the run time statistics is not well captured. Thus, we confirm that the ALP model better describes the motion of \textit{E. gracilis} under a weak light $0\ {\rm  W/m^2}<I\le 0.6\ {\rm  W/m^2}$, at least up to a length scale of $10^3\ \upmu$m.

Tab.~\ref{tab:euglena_results} shows the estimations of persistence length, swimming speed, and exponent estimated by DDM, and we compare the obtained exponent to the tracking results in Ref.~\cite{li2025biased}. The estimate of $\mu$ using DDM increases with light intensity, and the persistence length decreases accordingly, which is consistent with the finding in Ref.~\cite{li2025biased}. The ISFs of \textit{E. gracilis} are measured from a different biological replicate from that in Ref.~\cite{li2025biased}, which explains the quantitative difference between the measured $\mu$.

\begin{table}
    \centering
    \begin{tabular}{c||c|c|c|c}
    \hline\hline
        $I$ (W/m$^2$) & $\ell_p$ ($\upmu$m) & $v_0$ ($\upmu$m/s) & $\mu$ & $\mu$ from Ref.~\cite{li2025biased} \\ \hline
        0 & 3727 & 79.65 & 2.05 & 2.02 \\
        0.3 & 258 & 81.43 & 2.18 & 2.20 \\
        0.6 & 120 & 63.2 & 2.30 & 2.57 \\
        \hline\hline
    \end{tabular}
    \caption{Estimates of parameters of \textit{E. gracilis} with three light intensities $I$ from fitting the ISFs with the ALP model. The estimate of $\mu$ is compared with the published data in Ref.~\cite{li2025biased}.}
    \label{tab:euglena_results}
\end{table}
\section{Summary and discussion}\label{sec_summary}

In this article, we generalize the differential dynamic microscopy method to detect and characterize active L\'evy particles. We first analyze the asymptotic behavior of the intermediate scattering function of active L\'evy particles. We show that the ISF follows a scaling $k^{\mu-1}\tau$ over a wide range of $k$ with $2<\mu<3$, which differs qualitatively from the scaling $k^2\tau$ of classical run-and-tumble particles at small wave numbers. The oscillation of the ISF of ALPs persists at length scales of the order $10\ell_p$, if the speed variability is small, where the ISF of RTPs converges to an exponential function. This allows us to detect the difference between ALPs and RTPs in experiments. We show that a finite tumble time only rescales the effective time on large scales, given that the mean tumble time is finite. We analyze the effect of speed variability and show that it suppresses differences between ALPs and RTPs for $\sigma_v/\bar{v}\gtrsim 0.3$. To demonstrate an application of the ISF, we calculate the mean squared displacement of ALPs from its analytical ISF, which is exact for all time $t$ and exponent $\mu>2$. The exact analytical expression of MSD clearly shows the crossover between a ballistic $t^2$ regime and a superdiffusive $t^{4-\mu}$ regime for $2<\mu<3$.

Then we propose a protocol to measure and analyze the ISFs of active particles from experiments. To capture the asymptotic behavior of ISFs, data with multiple wavenumber $k$ ranging over lengths scales from $1\ell_p$ to $10\ell_p$ need to be fitted simultaneously. We test the protocol using synthetic image data and show that it can reliably detect ALPs and is robust under the particular range of length scales used in fitting. The method can extract the kinetic parameters within a 10\% relative error from the ground truth, if the speed variability of particles is moderate.

Finally, we apply our protocol to the experimental data of \textit{E. coli} and \textit{E. gracilis}. We find that the ALP model overfits the ISFs of \textit{E. coli}, the estimates of the exponent $\mu$ are unstable and can be unreasonably larger than 3. Instead, the ISFs of \textit{E. gracilis} are better fitted by the ALP model than the classic RTP model. These findings support the previous reports in Ref.~\cite{kurzthaler2024characterization} and~\cite{li2025biased}: \textit{E. coli} enters a diffusive regime on a length scale of the order $400\ \upmu$m, while \textit{E. gracilis} exhibits L\'evy walk at least up to a length scale of the order $10^3\ \upmu$m.

Our method provides a high-throughput alternative for detecting and characterizing active L\'evy particles. We highlight the importance of acquiring and analyzing data across scales. The characteristics of L\'evy walk are encoded in the asymptotic behavior of the intermediate scattering function of the particles, which can be captured only if ISFs of a wide range of wavenumbers are measured and analyzed simultaneously. The differential dynamic microscopy provides a versatile method for accessing data on a large length scale.

\begin{acknowledgments}
The authors thank Xiaqing Shi, Julien Tailleur, Hugues Chat\'e, and
Christina Kurzthaler for inspiring discussions. YZ acknowledges the support from the National Natural Science Foundation of China (No. 12304252). HPZ was supported by the National Key R\&D Program of China (grant no. 2021YFA0910700) and the National Natural Science Foundation of China (No. 12225410, No. 12074243).
\end{acknowledgments}

\bibliography{literature}

\end{document}